\documentclass[twoside,a4paper,11pt]{sea10}
\usepackage{graphicx}
\usepackage{hyperref}
\newcommand{\logg}  {$\log g$}

\begin{document}
\pagenumbering{arabic}
\pagestyle{myheadings}
\thispagestyle{empty}
{\flushleft\includegraphics[width=\textwidth,bb=58 650 590 680]{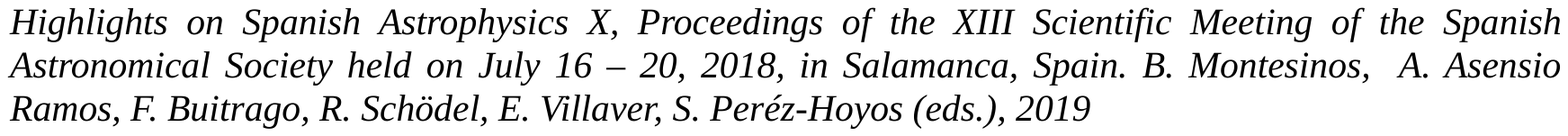}}
\vspace*{0.2cm}
\begin{flushleft}
{\bf {\LARGE
%
Can we really detect planets around evolved massive stars?
%
}\\
\vspace*{1cm}
%

E. Delgado Mena$^{1}$

%
}
\vspace*{0.5cm}
%
$^{1}$
Instituto de Astrof\'isica e Ci\^encias do Espa\c{c}o, Universidade do Porto, CAUP, Rua das Estrelas, 4150-762 Porto, Portugal\\
    
%
\end{flushleft}
%
\markboth{
Planets around evolved massive stars
}{ 
%
Delgado Mena et al.
%
}
\thispagestyle{empty}
\vspace*{0.4cm}
\begin{minipage}[l]{0.09\textwidth}
\ 
\end{minipage}
\begin{minipage}[r]{0.9\textwidth}
\vspace{1cm}
\section*{Abstract}{\small
%
The discovery of planets around massive stars is important for understanding how planet formation and evolution is conditioned by different stellar environments. However, current planetary search surveys have failed to detect planets around massive evolved stars. This lack of planets might be a consequence of the specifities of planet formation around such objects. Alternatively, the detection of planets around evolved massive stars might be hindered by the increasing stellar jitter as the stars evolve. In this project we target planets around evolved stars in open clusters, most of them with masses above 2\,M$_\odot$. We present the cases of three objects where long term (i.e. years) and high amplitude \textit{RV} signals of probably stellar origin are mimicking the presence of planets.
%
\normalsize}
\end{minipage}
%
%
%
\section{Introduction \label{intro}}
The last two decades have represented a strong success in the quest for exoplanets, mainly around late F and GKM stars. However, the search for planets around more massive stars (early F or A spectral types) is more difficult with the currently most used methods. For example, the Radial Velocity (\textit{RV}) technique, cannot be used in stars hotter than $\sim$\,6500\,K due to the increase in rotational velocities of those more massive stars and the lack of a sufficient number of spectral lines. Similarly, the larger radius of such stars hampers the discovery of planets with the transit method. Nevertheless, searching for planets around K giants, the evolved counterparts of those massive stars, has been a way to sort this problem out. The lower rotation rates and more crowded spectra of evolved stars facilitates the use of the \textit{RV} technique \cite{frink02,sato03,niedzielski15}.

\begin{figure}
\center
\includegraphics[width=8.5cm]{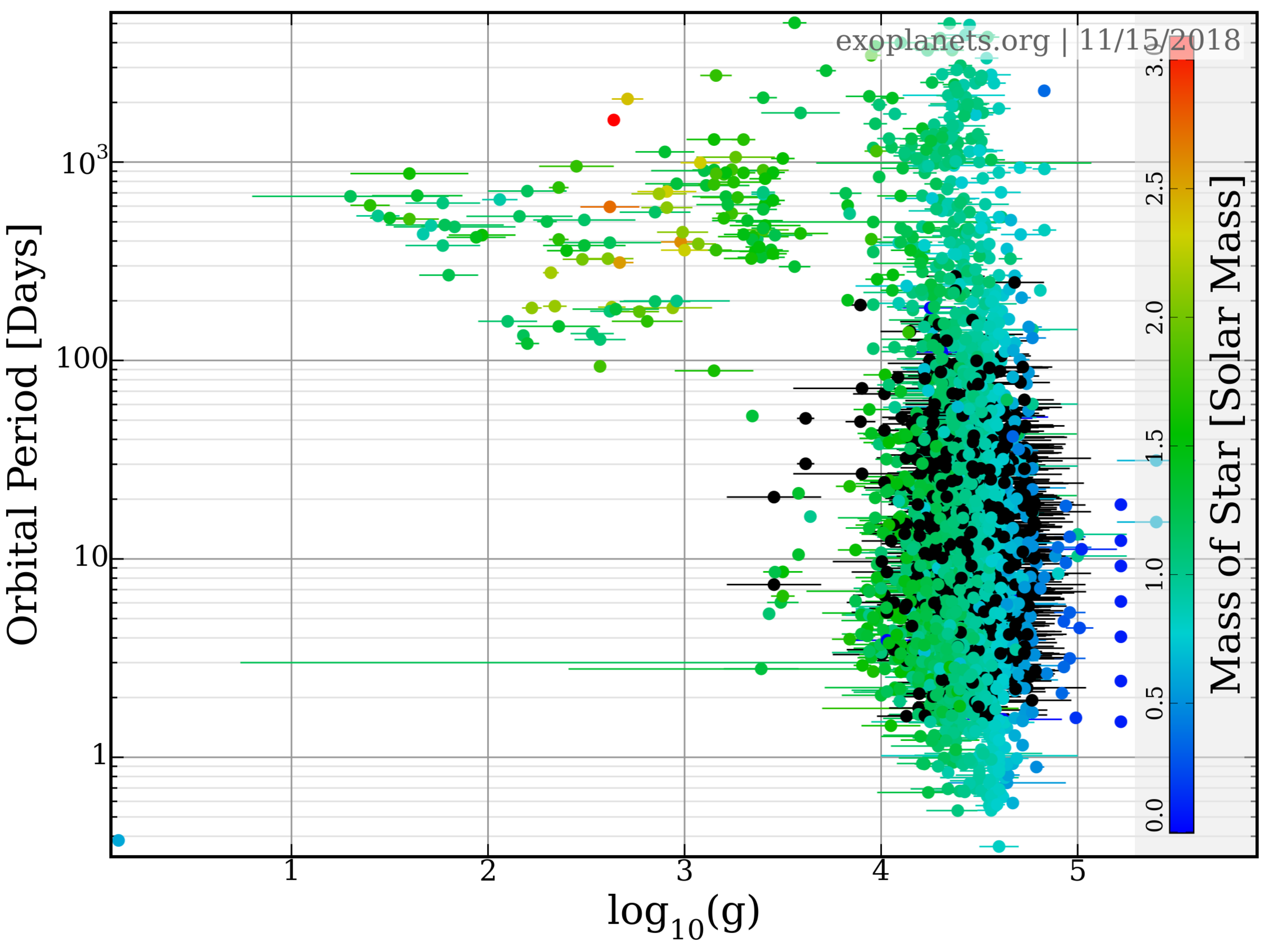} 
\caption{\label{period_logg} Orbital period of all planets discovered to date as a function of stellar surface gravity. Data and plot taken from exoplanets.org.}
\end{figure}

Still, current planetary search surveys have failed to detect close-in planets around massive evolved stars \cite{lillobox16}. In Fig. \ref{period_logg} we can see stars with \logg\,$<$\,3\,dex do not have planets with periods shorter than 100 days. Whether this lack of short period planets around massive stars has a primordial origin or is produced by planet engulfment is still debated \cite{villaver14,kunitomo11}. Moreover, several works have found that the frequency of massive planets is higher around more massive stars \cite{Johnson2010,Reffert2015} but with a maximum around 2\,M$_\odot$ and a sharp decrease for stars more massive than 2.5-3\,M$_\odot$. Indeed, long-term surveys of giant stars such as the Lick Observatory survey \cite{Reffert2015} or the EXPRESS survey \cite{Jones2016} have not found any planet in stars more massive than $\sim$2.7\,M$_\odot$. Certainly, an important concern when interpreting \textit{RV} variations in red giants is the presence of intrinsic stellar jitter (caused by p-mode oscillations) which shows a typical level of 10-20\,m\,s$^{-1}$ for stars at the base of the Red Giant Branch (RGB) \cite{Setiawan2004} and increases as the stars further evolve \cite{Hekker2008}. The modulation of active regions in red giants can produce large amplitude \textit{RV} variations as well. Therefore, analysing the stability of the \textit{RV} signals over several orbits (and for a time span larger than the stellar rotational period) is needed before claiming the presence of a planet. In this work we present new results for a survey around evolved stars in open clusters started by \cite{Lovis2007}, hereafter LM07. For more details about the present work we refer the reader to \cite{delgado18}.

\begin{figure}
\center
\includegraphics[width=15.5cm]{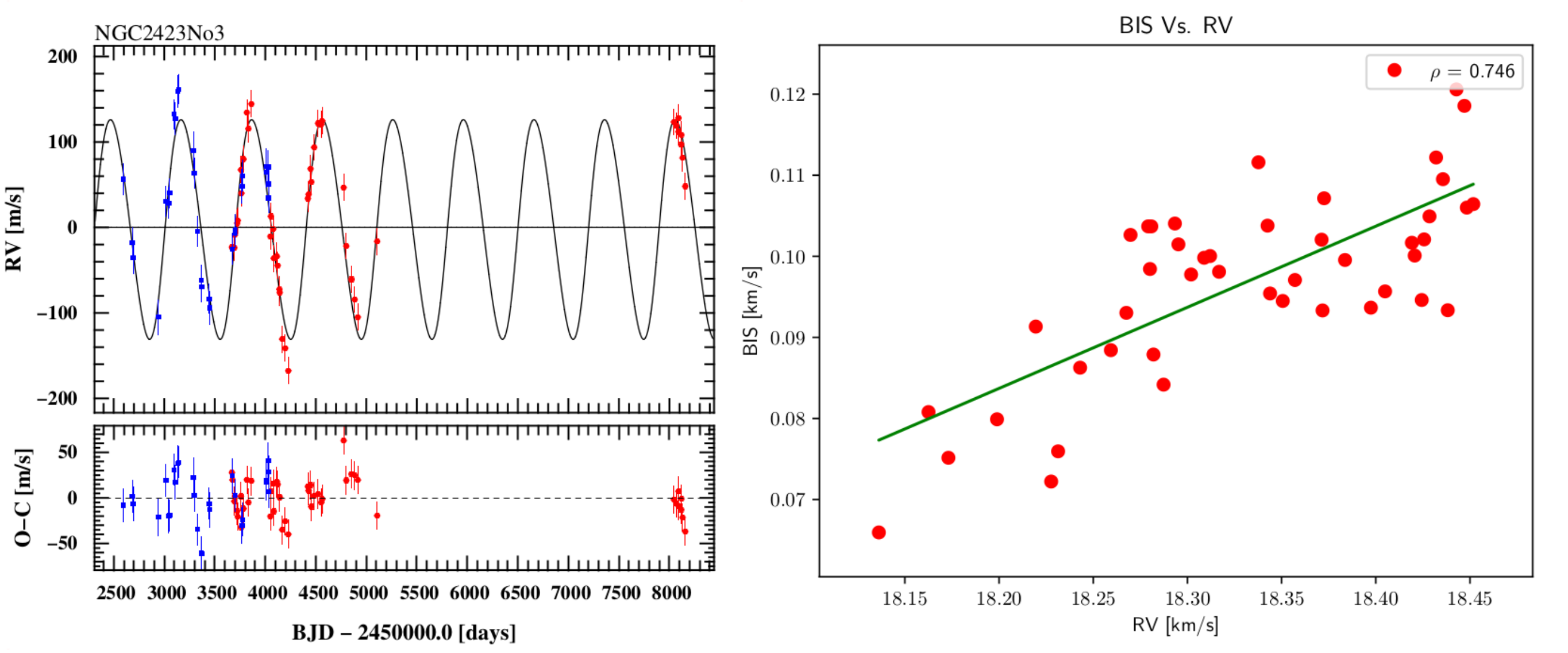} 
\caption{\label{NGC2423_bis} \textit{Left:} \textit{RV} curve as a function of time and as a function
of orbital phase for NGC2423No3. \textit{Right:} BIS vs \textit{RV} for NGC2423No3.}
\end{figure}

\section{NGC2423No3 and NGC4349No127}
In LM07, showing the first results of this survey, a planet in the cluster NGC2423 was announced after collecting 46 \textit{RV} points (28 with CORALIE and 18 with HARPS) during 1529 days. Additionally, a brown dwarf was discovered around NGC4349No127, by using 20 \textit{RV} measurements with HARPS along 784 days. In order to probe the planetary nature of the \textit{RV} signals, the Ca II H\&K lines and the Bisector Inverse Slope BIS \cite{Queloz2001} of the cross correlation function (CCF) were analyzed, but none of them showed a periodic variability. In the next two years after the publication of the results these two stars were followed up and additional data has been taken in 2017 and 2018. In the left panel of Fig. \ref{NGC2423_bis} we show the \textit{RV} variations of NGC2423No3 spanning 15 years of observations (although with a big gap in between 2009 and 2017) that can be fitted with a Keplerian fit with a period of 698 days. Considering the stellar mass (2.26$\pm$0.07\,M$_\odot$) of this object \cite{delgado16}, this signal would correspond to a planet with $m_{2}$ sin \textit{i}\,=\,9.6\,M$_{J}$ in a circular orbit with \textit{a}\,=\,2.02 AU. We note however, that the phase of the signal seems to slightly change along the time. The periodograms of the full-width-at-half-maximum (FWHM) of the cross-correlation function (CCF), the BIS and the H$\alpha$ line do not show a significant variation with the period of the \textit{RV}, however, we find that the BIS is strongly correlated with the \textit{RV} (see the right panel of Fig. \ref{NGC2423_bis}). This fact warns us about the possibility that the signals we are observing are related to inhomogeneities in the stellar surface and not with an orbiting body.

\begin{figure}
\center
\includegraphics[width=15.5cm]{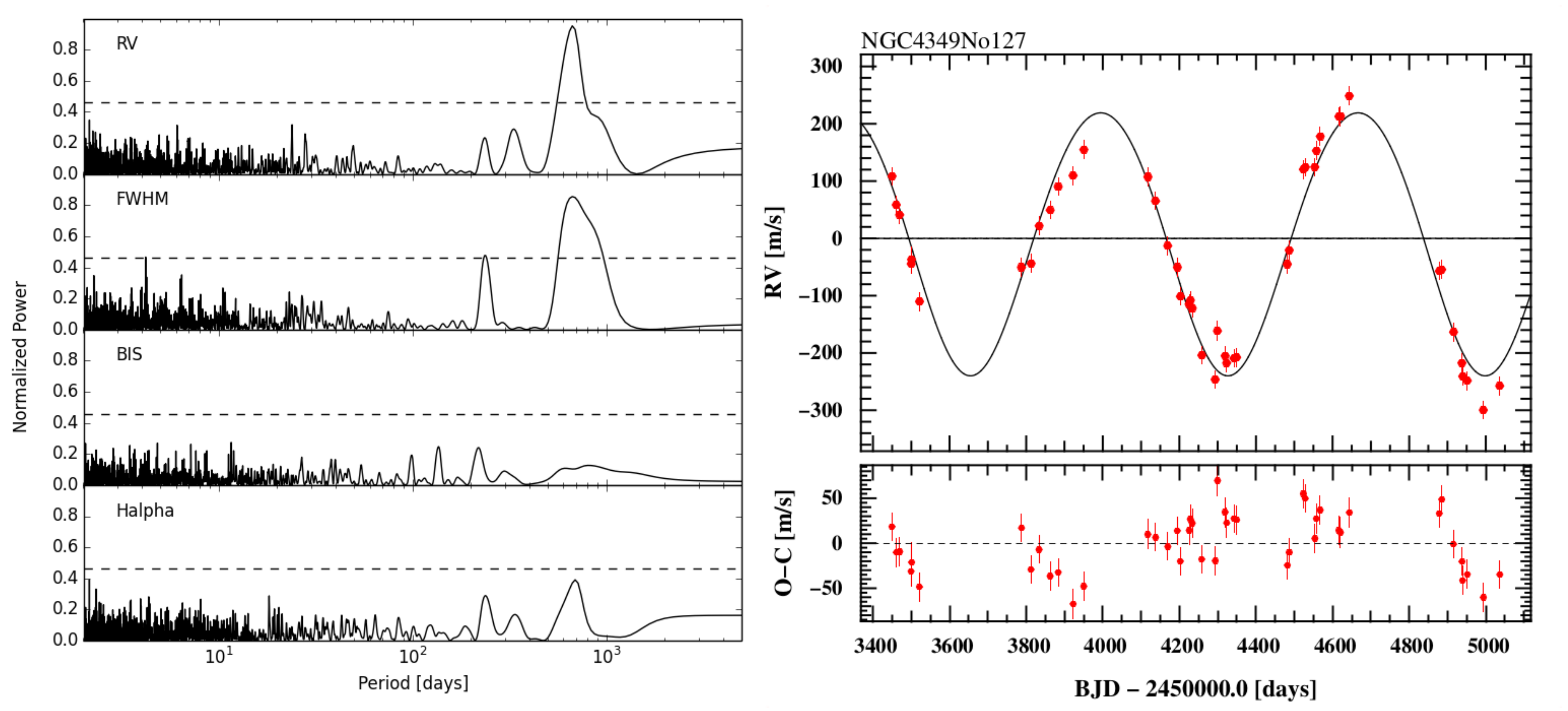} 
\caption{\label{NGC4349} \textit{Left:} Generalized Lomb-Scargle periodograms of \textit{RV}, FWHM, BIS and H$\alpha$ index
for NGC4349No127. The dashed line indicates the FAP at 1\% level. \textit{Right:} \textit{RV} curve as a function of time and as a function of orbital phase for NGC4349No127.}
\end{figure}

In the right panel of Fig. \ref{NGC4349} we show the \textit{RV} variations of NGC4349No127 for which 46 measurements were collected during 1587 days. The data can be fitted with a Keplerian orbit of 672 days, corresponding to a brown dwarf of $m_{2}$ sin \textit{i}\,=\,24.1\,M$_{J}$ (the stellar mass is 3.81$\pm$0.23\,M$_\odot$) in a near circular orbit. However, the periodogram of the FWHM has a strong peak at $\sim$666 days, the same period as observed in the \textit{RV} periodogram (see left panel of Fig. \ref{NGC4349}). This fact indicates that there are variations in the star's atmosphere with the exact same period as the \textit{RV}. Moreover, the H$\alpha$ periodogram also shows a signal at a similar period, with $P$\,$\sim$\,689 days (however just below the FAP = 1\% level). Therefore, the strong \textit{RV} signal is probably due to rotationally modulated active regions or long-period oscillations, and not as a result of an orbiting body.

\section{IC4651No9122}
\begin{figure}
\center
\includegraphics[width=15.5cm]{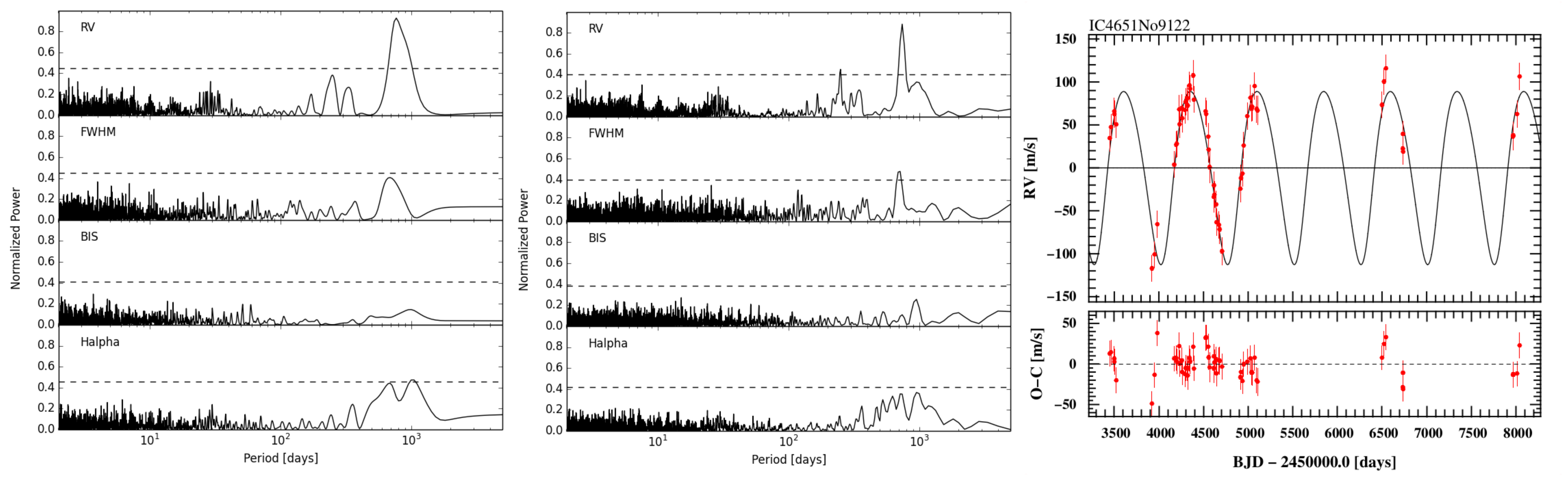} 
\caption{\label{IC4651} \textit{Left:} Generalized Lomb-Scargle periodograms of RV, FWHM and H$\alpha$ index
for IC4651No9122 with data taken until 2009. The dashed line indicates the FAP at 1\% level. \textit{Middle:} The same for the full dataset. \textit{Right:} \textit{RV} curve as a function of time and as a function of orbital phase for IC4651No9122.}
\end{figure}

In Fig. \ref{IC4651} we present the data for IC4651No9122, a star with a mass of 2.06$\pm$0.09\,M$_\odot$. The first set of data collected till 2009 (47 \textit{RV} points during $\sim$4.5 years) shows a very clear and significant peak at 771 days (see left panel in Fig. \ref{IC4651}). This \textit{RV} variation can be explained by the presence of a planet with $m_{2}$ sin \textit{i}\,=\,6.9\,M$_{J}$ in a 2.09 AU semi-major orbit. The H$\alpha$ periodogram shows a statistically significant signal at 1052 days and a peak below the FAP = 1\% line at 689.8 days. Moreover, the FWHM periodogram also shows a long-period signal at $\sim$689.8 days, close to the FAP = 1\% line. This signal is probably due to rotational modulation of active regions in the atmosphere. Although the most plausible explanation for the \textit{RV} variability is the presence of a planet, we considered a bit suspicious the fact that the period of the FWHM lies so close to the period of the planet candidate which in turn matches one of the peaks of the H$\alpha$ index periodogram. Therefore, we decided to re-observe this star during 2017 and 2018 and we also collected 6 additional points from the ESO archive. The periodogram of the complete dataset is shown in the middle panel of Fig. \ref{IC4651}. The \textit{RV} has now a peak at 741 days whereas the FWHM presents a significant peak, just above the FAP level, at 714 days, closer to the \textit{RV} period than with the initial set of data. The H$\alpha$ index also shows two peaks at 952 and 714 days but they are not statistically significant. The \textit{RV} curve with a Keplerian fit is shown in the right panel of Fig. \ref{IC4651}. Although with the current data we cannot rule out the presence of a planet around IC4651No9122, these results cast doubts on the planetary nature of the signal and more data will be needed to confirm the presence of a planet.

\section{Conclusions}

Long-period \textit{RV} variations with hundreds of days have been known to exist in several giant stars \cite{Walker1989,Larson1993,Hatzes1993} with \textit{RV} amplitudes in the order of $\sim$50-400\,m\,s$^{-1}$ which were attributed to rotationally modulated active regions \cite{Larson1993,Lambert1987} or radial and non-radial pulsations \cite{Hatzes1999}. Moreover, a new kind of pulsations with period of hundreds of days have been recently proposed to be manifestations of oscillatory convective modes \cite{Saio2015}. Although these pulsations are expected to be only present in high luminosity stars (log\,(L/L$_{\odot})$ $>$\,3\,dex for a 2\,M$_{\odot}$ star) they might be an explanation for the variability found in NGC4349No127, the most evolved of our targets \cite{delgado18}. Recently, this kind of pulsations have also been considered as a possible explanation for the \textit{RV} variations found in $\gamma$ Draconis with a semi-amplitude of 148\,m s$^{-1}$ and a period of 702 days that changes in phase and amplitude \cite{Hatzes2018}. Curiously, the suspicious \textit{RV} variations of our targets also have a period close to 700 days. Furthermore, there are other cases in the literature with long-period \textit{RV} variations in M giants that correlate with stellar activity indicators such as BIS or H$\alpha$ \cite{Lee2016,Bang2018}.

The three examples presented here clearly expose the difficulty of detecting planets around evolved stars, especially the most massive ones. Therefore, it is important to carry out long-term observations covering more than one period of the planet candidates (and for a time span larger than the stellar rotational period), to evaluate the stability of the hypothetical planetary signal and its possible relation with the rotational period of the star.

%
%
\small  
%
\section*{Acknowledgements}   
%
E.D.M. acknowledges the support from Funda\c{c}\~ao para a Ci\^encia e a Tecnologia (FCT) through national funds and from FEDER through COMPETE2020 by the following grants: UID/FIS/04434/2013 \& POCI--01--0145-FEDER--007672, PTDC/FIS-AST/1526/2014 \& POCI--01--0145-FEDER--016886, PTDC/FIS-AST/7073/2014 \& POCI-01-0145-FEDER-016880 and POCI-01-0145-FEDER-028953. E.D.M. acknowledges the support from FCT through Investigador FCT contract IF/00849/2015/CP1273/CT0003 and in the form of an exploratory project with the same reference.

%

%
\end{document}